\documentclass[aps, 10pt, prd, twocolumn, superscriptaddress, nofootinbib, showpacs=false, floatfix]{revtex4-2}
\usepackage{aas_macros}
\usepackage{amsmath, amssymb, amsfonts}
\usepackage{physics, tensor, natbib}
\usepackage{bm, physunits, soul}
\usepackage{xcolor}
\usepackage[colorlinks=true,linkcolor=blue,citecolor=blue,urlcolor=blue]{hyperref}
\newcommand{\lagr}{\mathcal{L}}
\newcommand{\amp}{\mathcal{A}}
\newcommand{\ham}{\mathcal{H}}
\newcommand{\ep}{\epsilon}
\newcommand{\cep}{\varepsilon}
\newcommand{\kap}{\kappa}
\newcommand{\mcal}{\mathcal{M}}
\newcommand{\realn}{\mathbb{R}}
\newcommand{\cd}{\mathcal{D}}

\newcommand{\nt}{\tilde{N}}
\newcommand{\tord}{\mathcal{T}}
\newcommand{\pow}{\mathcal{P}}
\newcommand{\p}{\partial}
\newcommand{\vol}{\mathcal{V}}
\newcommand{\re}{\operatorname{Re}}

\newcommand{\bx}{\vb{x}}
\newcommand{\br}{\vb{r}}

\def\lp{\ell_{\rm Pl}}
\newcommand{\ycs}{Y_{\rm CS}}
\newcommand{\tcs}{T_{\rm CS}}
\def\Vol{\operatorname{Vol}}
\newcommand{\ai}{\mathrm{Ai}}
\newcommand{\bi}{\mathrm{Bi}}

\begin{document}

\title{The Cosmological Constant as a Clock and the Vilenkin Wavefunction as a Resonance}

\author{Stephon Alexander}
\email{stephon\_alexander@brown.edu}
\affiliation{Brown Center for Theoretical Physics and Innovation, Brown University, Providence, Rhode Island 02912, USA}
\affiliation{Department of Physics, Brown University, Providence, Rhode Island 02912, USA}

\author{Marcell Howard}
\email{marcell\_howard@brown.edu}
\affiliation{Brown Center for Theoretical Physics and Innovation, Brown University, Providence, Rhode Island 02912, USA}
\affiliation{Department of Physics, Brown University, Providence, Rhode Island 02912, USA}

\begin{abstract}
We address the problem of time in quantum gravity using a covariant formulation of unimodular gravity. As a result, the cosmological constant is canonically conjugate to a global four-volume variable, which serves as a clock and leads to a time-dependent Wheeler--DeWitt equation. We construct exact, background-independent solutions related to the full Chern--Simons--Kodama state and demonstrate unitary evolution in minisuperspace. The Chern--Simons functional also provides an intrinsic gravitational clock. These solutions reveal that the Vilenkin wavefunction is a gravitational Gamow resonance, in which its non-normalizability follows from the outgoing boundary condition. The tunneling probability into the expanding branch increases with the positive cosmological constant. These results give a dynamical interpretation of the wavefunction of the universe and clarify the role of its boundary conditions.

\end{abstract}

\maketitle

\section{Introduction}
\label{sec:intro}

The problem of time has been a longstanding challenge in quantum gravity, reflecting a fundamental tension between general relativity and quantum mechanics~\cite{Isham92, Feinberg95, Kuchar11, Anderson17}. In the Hamiltonian formulation of general relativity, the lapse function acts as a Lagrange multiplier enforcing the Hamiltonian constraint. Upon quantization, this constraint becomes the Wheeler--DeWitt equation, which requires the wavefunction of the Universe to be annihilated by the Hamiltonian operator. This presents a conceptual difficulty because, in ordinary quantum mechanics, the Hamiltonian generates evolution through the Schr\"odinger equation. General relativity, however, is invariant under time reparametrizations, so the coordinate time $t$ is merely an arbitrary label rather than a physical observable. Consequently, the Hamiltonian constraint does not generate evolution with respect to an external physical time. Instead, the Wheeler--DeWitt equation describes a wavefunction of the entire Universe that appears frozen, with no explicit time evolution. Reconciling this timeless description with the dynamical Universe we observe lies at the heart of the problem of time in quantum gravity.

To state the issue simply, the usual time variable is wholly inappropriate for describing the evolution of a system where the symmetry is diffeomorphism invariance. This has inspired attempts to introduce a time variable that could be used in a more covariant fashion. Unimodular gravity supplies exactly that variable~\cite{Unruh89, Smolin09, Yamashita21, Alvarez23}. Unimodular gravity replaces the metric $g_{\mu\nu}$ with a symmetric, rank-2 object $\bar{g}_{\mu\nu}$ such that $\det\bar{g}_{\mu\nu} = 1$. The equations of motion for this new tensor are the traceless Einstein field equations, however the cosmological constant enters in as a constant of integration as opposed to a coupling constant at the level of the action. As a result, this formalism has been developed as a way to tackle the cosmological constant problem~\cite{Henneaux89, Ahmed04, Smolin09, Martin12, Padilla15, Alvarez23}. A covariant formulation of unimodular gravity which still respects diffeomorphism invariance was introduced in~\cite{Henneaux89}. In it, instead of somewhat arbitrarily setting $\det\bar{g}_{\mu\nu}$ to some constant function, we can instead introduce the cosmological constant as a spacetime-dependent field.

Our work is motivated in particular by the seminal analysis of Unruh and Wald~\cite{UnruhWald}, who recognized that the cosmological constant, when treated as an integration constant rather than a fixed parameter of the gravitational action, opens a possible resolution of the problem of time in quantum gravity. In this framework, the cosmological constant is canonically conjugate to a variable measuring spacetime four-volume, which can serve as a physical clock and restore a Schr\"odinger-like description of quantum evolution. This observation raises a fundamental question: can such an evolution be realized covariantly in the full theory of quantum gravity, beyond the symmetry-reduced setting of minisuperspace? Here we pursue this possibility through the generally covariant Henneaux--Teitelboim formulation of unimodular gravity, using the self-dual Ashtekar variables to construct exact, background-independent solutions from the Kodama state. In doing so, we provide a concrete covariant realization of the mechanism anticipated by~\cite{Unruh89}, while uncovering a new dynamical interpretation of the Vilenkin wavefunction as a quantum cosmological resonance.

These ideas have been developed even further over years, albeit in the restricted setting of minisuperspace (i.e. the space of homogeneous and isotropic metrics)~\cite{Magueijo11, Magueijo18, Magueijo20a, Magueijo20b, Magueijo21a, Magueijo22, Magueijo23, Alexandre22, Alexandre23a, Alexandre23b}. However, any true state in quantum gravity should be background independence. The need for a background independent state in quantum gravity has lead many to study the Kodama state~\cite{Kodama90}. 

In the self-dual Ashtekar-Barbero variables~\cite{Ashtekar86} with a fixed cosmological constant $\Lambda > 0$, the wave function constructed from the Chern-Simons functional of the connection, $\ycs[A]$, is an exact solution of every quantum constraint of general relativity (and we give a proof of this in Appendix~\ref{app:kodama_constraints}). It is one of very few exact, nonperturbative states of quantum gravity that are known. It has a semiclassical interpretation as quantum de Sitter spacetime~\cite{Smolin02}, and in minisuperspace it is the Fourier dual of the Hartle-Hawking and Vilenkin wavefunctions upon appropriate choices of contour~\cite{Magueijo20a, Alexander21}. Yet the Kodama state has long been regarded with suspicion, for reasons that are worth listing carefully.

First, it is a single state. For each value of $\Lambda$, there is exactly one Kodama state, so it does not come with a Hilbert space and hence has no dynamics. It is the problem of time in its starkest form: a wave function of the Universe that describes a whole de Sitter history at once and gives no way to ask what happens next. Second, $\Lambda$ enters as a fixed external parameter, and it does so non-analytically, through $1/\Lambda$ in the exponent. The cosmological constant is treated as a coupling even though the state that depends on it describes the entire Universe. Third, it is not normalizable in the naive inner product. This was emphasized in close analogy with the Chern-Simons state of Yang-Mills theory, which fails to be normalizable and whose perturbations include modes of negative energy~\cite{Witten03}. It was further argued that the Kodama state is not invariant under CPT. 

Progress has been made on the last point by constructing an inner product from the reality conditions on the Ashtekar connection~\cite{Alexander22}, under which the linearized state can be normalizable~\cite{Alexander25}, and by recognizing that for a complex Hamiltonian the relevant replacement for hermiticity may be $\pow\tord$ symmetry in the sense of Bender~\cite{Bender98, Bender02, Mostafazadeh02, Bender07, Bender10}. However, the first two points have received considerably less attention.

The central observation of this paper is that the first two objections are two faces of the same problem, and that the problem disappears once the cosmological constant is allowed to be a dynamical variable. To see why, it helps to recall how an energy eigenstate behaves in ordinary quantum mechanics. A stationary state $\phi_E(x)$ carries no information about time as time dependence sits entirely in a phase, and in a continuum of energies the eigenstates are only delta function normalizable. Physical states are then described as a superposition of wave packets. The Kodama state has every feature (and we will demonstrate this below) of such an eigenstate.

Our central result is a covariant realization of quantum evolution in the full self-dual theory: the Henneaux--Teitelboim four-volume clock turns the Hamiltonian constraint into a covariant Schr\"odinger equation, and a time-dependent extension of the full Kodama state supplies exact formal solutions beyond minisuperspace. In particular, at fixed $\Lambda$ the solution is $\Psi_\Lambda[A,T] = e^{i\Lambda T/\lp^2}\psi_\Lambda[A]$, with $\psi_\Lambda[A]$ being the Kodama functional of the background independent connection . This resolves the frozen-state aspect of the problem of time within the chosen unimodular framework, while leaving the construction of a full physical inner product and local relational observables as separate questions.

A further consequence of introducing unimodular time is a new interpretation of the Vilenkin wavefunction of the Universe. The purely outgoing tunneling prescription poses a familiar challenge for a unitary description: for a  real spectral parameter, a self-adjoint boundary condition does not admit a stationary solution with outgoing flux alone. In minisuperspace , we show that the same outgoing Airy wave instead defines a Gamow-type resonance~\cite{Gamow28, Bohm89, Moiseyev98, Civitarese04, delaMadrid05} through a pole of the analytically continued scattering amplitude. Its non-normalizability is then understood as a property of a generalized resonance mode,  as opposed to a failure of normalizability of the physical wave packets. Crucially, the unimodular clock gives the resonance width a direct dynamical meaning as an amplitude decay rate. This distinction suggests a new way to interpret the Vilenkin prescription without identifying it with the complete physical quantum state~\cite{Alexandre23b, Gielen26}.

Here, we build on the framework established in~\cite{Smolin11, Magueijo18} as well as the work of~\cite{Yamashita21, Yamashita26} but works at the self-dual value of the Barbero-Immirzi parameter, where the state is Kodama's original one and its de Sitter semiclassical limit is available. The aim is to carry the unimodular construction through in the full theory, without symmetry reduction, and to extract what the full theory adds.

The paper is organized as follows. In Section~\ref{sec:HT}, we review the Henneaux-Teitelboim prescription and show how it supplies a gauge invariant global time conjugate to $\Lambda$. In Section~\ref{sec:canonical}, we carry out the canonical analysis in self-dual Ashtekar variables and identify the gauge invariant pair $(T[\Sigma], \Lambda)$. In Section~\ref{sec:quantization}, we quantize and obtain a Schr\"odinger equation in unimodular time. Section~\ref{sec:kodama}, shows that the Kodama state is a stationary state of this equation. In Section~\ref{sec:CStime}, we show that the Chern-Simons functional defines a clock that tracks unimodular time, and in Section~\ref{sec:wavepackets} we construct wave packets sharply peaked on Chern-Simons time. Section~\ref{sec:minisuperspace}, treats the minisuperspace limit, where we establish orthogonality of the Kodama states and conservation of the norm. Section~\ref{sec:vilenkin_res} develops the outgoing Vilenkin solution as a Gamow-type resonance and identifies its width in unimodular time. We close in Section~\ref{sec:discussion} by discussing the resulting interpretation of the wavefunction of the Universe and the open questions.

\paragraph*{Conventions}

We use the mostly plus metric signature, i.e. $\eta_{\mu\nu} = (-,+,+,+)$ and units where $c = 1$. We use boldface letters $\bx$ to indicate 3-vectors and we use $x$ and $p$ to denote 4-vectors. Conventions for the curvature tensors, covariant and Lie derivatives are all taken from Carroll \cite{Carroll04}. Greek indices $(\mu, \nu, \ldots)$ and Latin indices $(a, b, c,\ldots)$ denote spacetime indices and Latin indices $(i, j, k, \ldots)$ denote indices for the internal space. $\ep^{\mu\nu\rho\sigma}$ and $\ep_{\mu\nu\rho\sigma}$ are the Levi Civita densities with $\ep^{0123} = \ep_{0123} = 1$. 

\section{The Henneaux-Teitelboim Prescription}\label{sec:HT}

The main idea for this work starts from the cosmological constant, $\Lambda$ being the canonical partner to the 4-volume. We start from the Holst action~\cite{Holst96} with a cosmological constant and add the Henneaux-Teitelboim term~\cite{Henneaux89},

\begin{equation}
	\begin{split}
		S = {}&\frac{1}{4\kap}\int\ep_{IJKL}e^I\wedge e^J\wedge R^{KL} - \frac{1}{2\kap\gamma}\int e^I\wedge e^J\wedge R_{IJ}\\ &- \frac{1}{4!\kap}\int\Lambda\ep_{IJKL}e^I\wedge e^J\wedge e^K\wedge e^L
		+ \frac{1}{\kap}\int\Lambda\dd\tord,
	\end{split}
\end{equation}
here $\kap = 8\pi G$, $\{e^I\}$ are the cotetrad fields, $R^{IJ} = \dd\omega^{IJ} + \omega^{I}{}_{K}\wedge\omega^{KJ}$ is the curvature of the Lorentz connection, $\gamma$ is the Barbero-Immirzi parameter, $\Lambda = \Lambda(\br, t)$ is a spacetime-dependent field and $\tord$ is a three-form. In four spacetime dimensions, a three-form can be represented as the following
\begin{equation}
	\tord = \frac{1}{3!}\tord_{\nu\lambda\rho}\dd{x^\nu}\wedge\dd{x^\lambda}\wedge\dd{x^\rho}.
\end{equation}
Again, a consequence of working in four spacetime dimensions is that the dual object to a three-form is a \emph{vector density} that we call $T^\mu = T^\mu(\br, t)$, which is related to the original three-form by
\begin{equation}
	T^\mu = \frac{1}{3!}\ep^{\mu\nu\lambda\rho}\tord_{\nu\lambda\rho}, \hspace{0.5cm}\tord_{\nu\lambda\rho} = \ep_{\mu\nu\lambda\rho}T^\mu.
\end{equation}
The exterior derivative of a three-form, $\dd\tord$ to get a four-form whose dual is a scalar function:
\begin{equation}
	\dd\tord = \p_\mu T^\mu \dd[4]{x}.
\end{equation}
Recall the four-volume can be represented as 
\begin{equation}
	\frac{1}{4!}\ep_{IJKL}e^I\wedge e^J\wedge e^K\wedge e^L = \sqrt{-g}\dd[4]{x}, 
\end{equation}
and thus we can represent the last two terms of the modified Holst action as

\begin{equation}
    \begin{split}
        S_{\rm HT} &= \frac{1}{\kap}\int\Lambda\dd\tord - \frac{1}{4!\kap}\int\Lambda\ep_{IJKL}e^I\wedge e^J\wedge e^K\wedge e^L\\ &= \frac{1}{\kap}\int\Lambda(\p_\mu T^\mu - \sqrt{-g})\dd[4]{x}.
    \end{split}
\end{equation}
Consequently, both $\Lambda$ and $T^\mu$ act as Lagrange multipliers. If we vary the integrand with respect to $\Lambda$ and the vector density respectively, we find
\begin{equation}
	\p_\mu T^\mu = \sqrt{-g}, \quad \p_\mu\Lambda = 0.
\end{equation}
This tells us that the the spacetime-dependent function, $\Lambda(\br)$, in the Henneaux-Teitelboim action is constrained to be a constant and can be taken out of the integral. This is important because we can now define the \emph{unimodular time} $T$ for the spacelike hypersurface $\Sigma$ given by\footnote{The 0-th component is all that's left because it is the only variable that's left in the object $T^\mu$ because it's the only component with a time derivative and is hence dynamical.}

\begin{equation}
	T[\Sigma] = \int\dd\tord = \int_{\Sigma}\tord = \int_{\Sigma}\dd[3]{r}T^0(\br, t).
\end{equation}
The unimodular time is gauge invariant under $\tord\rightarrow\tord + \dd\theta$ where we assume $\theta$ is a 2-form that goes to zero at the boundary. This gauge symmetry means we can define the unimodular time up to an overall constant, consistent with its interpretation of being a global clock.

We've employed this construction from \cite{Henneaux89} because we are explicitly looking for a covariant formulation of the Wheeler-DeWitt equation.

\section{Canonical Analysis}\label{sec:canonical}

Now we switch to the Ashtekar variables as to make the Hamiltonian constraint easier to solve. We do the following procedure: first work in usual 3+1 decomposition of spacetime where we foliate our hypersurfaces of constant $t$~\cite{Arnowitt59, Arnowitt08}. Then we switch to using the densitized triads $E^a_i = \sqrt{q}e^a_i$, where $q = \det(q_{ab})$ is the determinant of the induced metric on the spatial manifold. Next we work in the connection representation with $A^i_a = \Gamma^i_a + iK^i_a$ with $\Gamma$ being the spin connection and $K^i_a$ is the extrinsic curvature and we've set the Barbero-Immurzi parameter $\gamma = i$. The action that we started with becomes

\begin{equation}
    \begin{split}
        S = \int \dd{t}\int_\Sigma \dd[3]{r}\,\Big[&\frac{1}{i\kap}E^a_i\dot A^i_a + \frac{1}{\kap}\Lambda\,\dot T^0 - A^i_0G_i - N^aV_a\\ 
    &- \nt\,\ham - \frac{1}{\kap}T^a\p_a\Lambda\Big],
    \end{split}
    \label{eq:canonical}
\end{equation}
where overhead dots correspond to the Lie derivative in the direction of the vector field $\p_t$, $\ham$, $V_a$, $G_i$ are the scalar, vector, and Gauss constraint respectively, with Lagrange multipliers $N$, $N^a$, and $A^i_0$ with $\tilde{N} = N/E$. We also used integration by parts of the $\Lambda\p_a T^a$ term. The scalar constraint is given by

\begin{equation}
	\begin{split}
		\ham &= \frac{1}{2\kap}\qty(\ep_{ijk}E^a_iE^b_jF^k_{ab} + \frac{\Lambda}{3}\ep_{ijk}\ep_{abc}E^a_iE^b_jE^c_k)\\ &= \frac{1}{2\kap}\ep_{ijk}E^a_iE^b_jF^k_{ab} + \frac{\Lambda}{\kap}\det E,
	\end{split}
    \label{eq:hamiltonian}
\end{equation}
where the $F^i_{ab} = \p_aA^i_b - \p_b A^i_a + \ep_{ijk}A^j_aA^k_b$ is the field-strength tensor for the gauge field. In terms of canonical variables, the other constraints are

\begin{equation}
	V_a = E^b_jF^j_{ab}, \hspace{0.5cm}G_i = \cd_aE^a_i,
    \label{eq:other_constraints}
\end{equation}
where $\cd_a$ is the covariant derivative with connection $A^i_a$. Adding in the Henneaux-Teitelboim term gives us an additional set of Poisson brackets

\begin{equation}
	\begin{split}
	    \pb{A^i_a(\br)}{E^b_j(\br')} &= i\kap\delta^i_j\delta^b_a\delta^{(3)}(\br - \br'),\\ \pb{T^0(\br)}{\Lambda(\br')} &= \kap\delta^{(3)}(\br - \br'),
        \label{eq:poisson_brackets}
	\end{split}
\end{equation}
whereas the components of $T^a$ enforces the constraint

\begin{equation}
	\p_a\Lambda = 0.
\end{equation}

Since there is no $T^0$ term in $\ham$, we can very quickly conclude that this constraint commutes with $\ham$. Since $\p_a\Lambda$ is a vector, we know it will transform covariantly under spatial diffeomorphisms. Because the vector constraint only moves the gravitational variables, we can calculate the Poisson bracket between the gauge field and the smeared shift/gauge parameter $\xi_a$

\begin{equation}
	\begin{split}
		\pb{A^i_a(\br)}{\int\xi^cV_c} = -\xi^bF^i_{ab}(\br), 
	\end{split}
\end{equation}
which is the Lie derivative of $A^i_a$ along the direction of the gauge parameter $\xi^a$ which we write $\lagr_\xi A^i_a = -\xi^bF^i_{ab}$. The Lie derivative to the scalar function $\Lambda$ and the vector density $T^0$ can be found in Carroll's textbook.

\begin{equation}
	\lagr_\xi\Lambda = \xi^a\p_a\Lambda, \hspace{0.5cm}\lagr_\xi T^0 = \p_a(\xi^a T^0).
\end{equation}
Note, these relations could've been generated by the object $D[\xi]$ defined as

\begin{equation}
	D[\xi] \equiv -\frac{1}{\kap}\int_{\Sigma}\dd[3]{r}\xi^a T^0\p_a\Lambda,
\end{equation}
via the Poisson brackets

\begin{equation}
	\pb{\Lambda}{D[\xi]} = \xi^a\p_a\Lambda, \hspace{0.5cm}\pb{T^0}{D[\xi]} = \p_a(\xi^aT^0).
\end{equation}

Because $T^0$ transforms under a spatial diffeomorphism by $-\kap\int\p_a\xi^a$ (this comes from integration by parts of the $\Lambda$ term). Therefore, the \emph{only} gauge invariant object that we can construct from $T^0$ is the global time

\begin{equation}
	T[\Sigma] = \int_{\Sigma}\dd[3]{r}T^0(\br, t).
\end{equation}
Additionally, the only gauge invariant object that can be constructed from $\Lambda$ is just its constant value. Therefore, we have the following gauge invariant canonical pair

\begin{equation}
	\pb{T[\Sigma]}{\Lambda} = \kap.
\end{equation}

We can work out the Hamilton's equations for this conjugate pair. Since $\Lambda$ is a constant, $\dot\Lambda = 0$. However, instead of computing $\dot{T}[\Sigma]$ directly, it'll be easier to compute $\dot{T}^0$

\begin{equation}
	\dot{T}^0(\br, t) = N\sqrt{q},
\end{equation}
which further implies that

\begin{equation}
	\dot{T}[\Sigma] = \int_{\Sigma}\dd[3]{r}\dot{T}^0(\br,t) = \int_{\Sigma}\dd[3]{r}N\sqrt{q}.
\end{equation}
if we integrate with respect to $t$ and recall that we foliated the manifold, $\mcal$, by $\mcal = \realn\times\Sigma$, then we can express the unimodular time as 

\begin{equation}
    T[\Sigma] = \int_{\Sigma}\dd[3]{r}\int_{\realn}\dd{t}N\sqrt{q} = {\rm Vol}(\mcal).
    \label{eq:manifold_clock}
\end{equation}
So we see that the global time is exactly given by the integral of the four-volume. This result will come in handy when we'll want to relate the Chern-Simons functional as a clock that the wave function $\Psi$ peaks at.

\section{A Covariant Schr\"odinger Equation}\label{sec:quantization}

We quantize in the connection representation for the gravitational variables and in the $T^0$ representation for the unimodular sector. We do the usual procedure where we promote our canonical variables to operators and replace the Poisson brackets with commutators $\pb{A}{E} \mapsto (i\hbar)^{-1}\comm{\hat A}{\hat E} = -\lp^2$ and $\comm{\hat T^0}{\hat \Lambda} = i\lp^2$. States are then functionals $\Psi[A,T^0]$, holomorphic in $A$, and the brackets are represented by

\begin{equation}
	\hat E^a_i(\br) = \lp^2\fdv{A^i_a(\br)}, \hspace{0.5cm}	\hat\Lambda(\br) = -i\lp^2\fdv{T^0(\br)}.
\end{equation}
We order the triads to the left of the curvature in the scalar constraint. Then $\hat\ham(\br)\Psi = 0$ reads

\begin{equation}
	2i\lp^2\det\hat E(\br)\fdv{\Psi}{T^0(\br)} = \ep_{ijk}\hat E^a_i\hat E^b_j\hat F^k_{ab}(\br)\Psi.
\end{equation}
It should be noted that we are assuming that $\hat\Lambda$ commutes with $\hat E$. We've demonstrated the equivalence of $T^0$ and $N\sqrt{q}$ and $\det E = \sqrt{q}$, so one can wonder whether this step is justified. We will assume it is for now and come back to this objection in the (near) future. The constraint from $\p_a\Lambda = 0$ implies

\begin{equation}
	\p_a\fdv{\Psi}{T^0} = 0,
\end{equation}
which constrains $\Psi$ to be independent of $\br$ which further implies that $\Psi$ depends on $T^0$ but solely through the global clock $T$. Thus, we can swap variables from $T^0$ to $T$. This further means we can write $\hat\Lambda = i\lp^2\p_T$. For a non-vanishing $\det \hat E$, we can thus write

\begin{equation}
	i\lp^2\,\pdv{\Psi[A,T]}{T} = \hat h(x)\Psi[A,T],
    \label{eq:covariant_Schrodinger}
\end{equation}
where

\begin{equation}
	\hat h(\br)\equiv\frac{1}{2\det\hat E}\,\ep_{ijk}\hat E^a_i\hat E^b_j\hat F^k_{ab}(\br).
\end{equation}
We can see that~\eqref{eq:covariant_Schrodinger} can really be thought of as a covariant formuation of the Schr\"odinger equation. This comes from the fact that $T$ is a gauge invariant variable that the wave function evolves with respect to. The equation itself is reminiscent of the Tomanaga-Schwinger equation~\cite{Tomonaga46, Schwinger48}, a formulation of quantum mechanics originally invented as a precursor of quantum electrodynamics in order to understand the dynamics of electrons.

The operator $\hat h(\br)$ has density weight zero. Since the left hand side does not depend on $\br$, physical states must obey $\hat h(\br)\Psi = \hat h(\br')\Psi$ $\forall \br, \br'$, which is the quantum statement that the value of $\Lambda$ inferred locally from the geometry is the same everywhere.

\section{The Kodama State as a Stationary State}\label{sec:kodama}

The Chern-Simons functional is

\begin{equation}
	\ycs[A] = \int_\Sigma \dd[3]{r}\,\ep^{abc}\Big(A^i_a\dd_b A^i_c + \frac13\ep_{ijk}A^i_aA^j_bA^k_c\Big),
\end{equation}
and its functional derivative is

\begin{equation}
	\fdv{\ycs}{A_{ia}} = \epsilon^{abc}F^i_{bc}.
\end{equation}

For a \emph{fixed eigenvalue} $\Lambda$ of $\hat\Lambda$ define\footnote{The succeeding arguments definitely hold if we demand that our Universe be an eigenstate of the $\hat\Lambda$ \cite{Henneaux89}.}

\begin{equation}
	\psi_\Lambda[A] = \exp(-\frac{3}{2\Lambda\lp^2}\,\ycs[A]),
    \label{eq:kodama_state}
\end{equation}
which is the wave function that's found from solving

\begin{equation}
	\qty(F^k_{ab} + \frac{\Lambda}{3}\ep_{abc}\hat E^c_k)\psi_\Lambda = 0
\end{equation}
If we move the term that's proportional to $\Lambda$ to the other side and contract both sides with a factor of $\ep_{ijk}E^a_iE^b_j$, we get

\begin{equation}
	\ep_{ijk}\hat E^a_i\hat E^b_j\hat F^k_{ab}\psi_\Lambda = -2\Lambda\det\hat E\,\psi_\Lambda \Leftrightarrow \hat h(\br)\,\psi_\Lambda = -\Lambda\,\psi_\Lambda,
\end{equation}
and therefore each $\psi_\Lambda$ is an eigenstate of the unimodular Hamiltonian and

\begin{equation}
	\Psi_\Lambda[A,T] = e^{i\Lambda T/\lp^2}\,\psi_\Lambda[A],
\end{equation}
is a stationary solution of the covariant Schr\"odinger equation we wrote down in Eqn.~\eqref{eq:covariant_Schrodinger}. The identification of $\psi_\Lambda$ as an eigenstate of the unimodular Hamiltonian can essentially be traced back to works such as~\cite{Smolin02} where it was argued that the Kodama state can be seen as a WKB wave functional whose semiclassical limit is de Sitter spacetime. Because $\ycs$ and $T$ are global quantities, $\Psi_\Lambda$ is invariant under small gauge transformations and spatial diffeomorphisms. The general solution to this Schr\"odinger equation can be written as the following superposition

\begin{equation}
	\Psi[A,T] = \int \dd\Lambda\,\amp(\Lambda)\,e^{i\Lambda T/\lp^2}\,\psi_\Lambda[A],
\end{equation}
with $\amp(\Lambda)$ an arbitrary amplitude, which we take to be supported on $\Lambda > 0$. This is essentially a realization of what was done in~\cite{Unruh89} where we have been able to identify the $\psi_\Lambda$ that enters into the general solution.

Given this particular construction, we can also say that \emph{for fixed eigenvalue} $\Lambda$, the state $\Psi[A, T]$ will also satisfy the other constraints. This comes from the fact that for fixed $\Lambda$, the operator $\hat E^a_i$ passes over the amplitude $\amp(\Lambda)\,e^{i\Lambda T/\lp^2}$ harmlessly while the Kodama state is simultaneously annihilated by all three constraints (and we give a brief proof of that in the Appendix~\ref{app:kodama_constraints}). 

\section{The Chern-Simons Functional as a Clock}\label{sec:CStime}

For this part of the calculation, we shall work in the gauge in which $N^a = T^a = 0$ and $A^i_0 = 0$, but we keep $N$ arbitrary. We can compute the time evolution of the gauge field using its Poisson bracket

\begin{equation}
	\begin{split}
		\dot{A}^i_a &= i\kap\pb{A^i_a}{\int\nt\ham}\\&= \frac{1}{2}i\nt(2\ep_{ijk}E^b_jF^k_{ab} + \Lambda\ep_{abc}\ep_{ijk}E^b_jE^c_k).
	\end{split}
\end{equation}

Using the self-dual constraint where $F^k_{ab} = -\frac{\Lambda}{3}\ep_{abc}E^{ck}$, the time evolution of the gauge field becomes

\begin{equation}
	\begin{split}
		\dot{A}^i_a &= \frac{1}{2}i\nt(2\ep_{ijk}E^b_jF^k_{ab} + \Lambda\ep_{abc}\ep_{ijk}E^b_jE^c_k)\\ &= \frac{i\nt}{2}\frac{\Lambda}{3}\ep_{ijk}\ep_{abc}E^b_jE^c_k.
	\end{split}
\end{equation}
Thus, it is easy to see that

\begin{equation}
	E^a_i\dot{A}^i_a = i\nt\Lambda\cdot \frac{1}{3!}\ep_{ijk}\ep_{abc}E^a_iE^b_jE^c_k = i\nt\Lambda\det E = i\Lambda N\sqrt{q}.
\end{equation}

Next, we relate the evolution of the Chern-Simons functional to the evolution of the unimodular time.

\begin{equation}
	\begin{split}
		\dot Y_{\rm CS} &= \pb{\ycs}{\int\nt\ham}\\ &= -\frac{2\Lambda}{3}\int\dd[3]{r}E^a_i\dot{A}^i_a = -\frac{2i\Lambda^2}{3}\dot{T},
	\end{split}
\end{equation}
where we recall that

\begin{equation}
	\dot T = \int_{\Sigma}\dd[3]{r}N\sqrt{q}.
\end{equation}

Because $\Lambda$, $N$, and $\sqrt{q}$ are all positive quantities, we can identify the following

\begin{equation}
	\dv{t}\Re\ycs = 0. 
\end{equation}
If we define the Chern-Simons time as 

\begin{equation}
	\tcs[A] \equiv -\frac{3}{2\Lambda^2}\Im\dot{Y}_{\rm CS} = \dot{T}, 
\end{equation}
and thus

\begin{equation}
	\dv{t}(T - \tcs[A]) = 0.
\end{equation}
When we reduce to minisuperspace, and in particular de Sitter spacetime, the Chern-Simons time becomes

\begin{equation}
	\tcs = \frac{3\dot{a}^3\vol_c}{\Lambda^2} =\frac{a^3\vol_c}{3H}.
\end{equation}
Taking all the previous results together, we've shown that the unimodular time $T[\Sigma]$ satisfies the first condition necessary to be a ``good clock": our dynamical time variable is decoupled from all other dynamical variables.

\section{Wave Packets Peaked On Chern-Simons Time}\label{sec:wavepackets}

Suppose we chose the Kernel in the inner product to be a wave packet

\begin{equation}
	\amp(\Lambda) = \exp(-\frac{(\Lambda - \Lambda_0)^2}{2\sigma^2}),
\end{equation}
with $\sigma \ll \Lambda_0$. If we write $\Lambda = \Lambda_0 + \delta$, and  expand $1/\Lambda$ with $\delta \ll \Lambda_0$ then the Kodama state becomes

\begin{equation}
	\begin{split}
	    \psi_\Lambda \simeq \exp(-\frac{3}{2\lp^2\Lambda_0}\qty(1 - \frac{\delta}{\Lambda_0})\ycs[A]).
	\end{split}
\end{equation}

Starting from the general solution to our covariant Schr\"odinger equation 

\begin{equation}
	\Psi[A,T] = \int\dd\Lambda\, \amp(\Lambda)\,e^{i\Lambda T/\lp^2}\,\psi_\Lambda[A],
\end{equation}
in the $\Lambda = \Lambda_0 + \delta$ expansion the wave function becomes

\begin{equation}
	\begin{split}
		\Psi[A,T] &= \Psi_{\Lambda_0}[A, T]\int\dd\delta e^{-\frac{\delta^2}{2\sigma^2}}e^{i\delta(T - \tau[A])/\lp^2},
	\end{split}
\end{equation}
where 

\begin{equation}
	\Psi_{\Lambda_0}[A, T] \simeq e^{i\Lambda_0T/\lp^2}\exp(-\frac{3}{2\Lambda_0^2\lp^2}\ycs[A]),
\end{equation}
with

\begin{equation}
	\begin{split}
	    \tau[A] = \frac{3i}{2\lp^2\Lambda_0^2}\ycs[A] &= \tcs[A] + iS[A],\\ S[A] &= \frac{3}{2\Lambda_0^2}\Re\ycs[A]
	\end{split}
\end{equation}
We thus have

\begin{equation}
    \Psi[A,T]\simeq\sqrt{2\pi}\,\sigma\,\Psi_{\Lambda_0}[A,T]\,\exp(-\frac{\sigma^2}{2\lp^4}\big(T -\tau[A]\big)^2),
\label{eq:packetresult}
\end{equation}
and we can write down the amplitude of the wave function

\begin{align}
    |\Psi[A, T]|\simeq{}&\sqrt{2\pi}\,\sigma\,\exp(-\frac{3\,\re\ycs}{2\Lambda_0\lp^2})\nonumber\\
&\times\exp\Big\{-\frac{\sigma^2}{2\lp^4}\Big[\big(T-\tcs[A]\big)^2 - S^2[A]\Big]\Big\}.
\label{eq:modpacket2}
\end{align}
From here, we can conclude that our dynamical time variable satisfies the second criterion for being a ``good clock": that the wave function (in this case $\Psi[A, T]$) be sharply peaked in $T$ around some monotonic function of the global time variable which we have shown that $\tcs[A]$ matches the description.

\section{The Minisuperspace Limit and Unitarity}\label{sec:minisuperspace}

Up until this point, everything we've been doing has been background independent. We shall break that here by moving to minisuperspace. In this limit, the connection and it's partner become $A^i_a = i\dot{a}\delta^i_a$, $E^a_i = a^2\delta^a_i$, and therefore their commutator is simply
\begin{equation}
    \comm{\hat{\dot{a}}}{\hat{a}^2} = i\frac{\lp^2}{3\vol_c}.
    \label{eq:gravity_commutator}
\end{equation}

Similarly, the commutator between the unimodular time and the cosmological constant is simply
\begin{equation}
    \comm{\hat{T}}{\hat{\Lambda}} = \frac{i\lp^2}{3\vol_c}.
    \label{eq:unimodular_commutator}
\end{equation}
In the minisuperspace limit, the Kodama state simplifies to

\begin{equation}
	\psi_\Lambda = \exp(\frac{3iu}{\Lambda\lp^2}),
    \label{eq:minisuperspace_kodama}
\end{equation}
where $u = \dot{a}^3\vol_c$. If we take the norm of this function using the usual inner product, we can show the following orthogonality condition on the states

\begin{equation}
	\begin{split}
		\int\dd{u}\psi_\Lambda^*\psi_{\Lambda'}  = \int\dd{u}\exp(-\frac{3i}{\lp^2}\qty(\frac{1}{\Lambda} - \frac{1}{\Lambda'})u).
	\end{split}
\end{equation}
we can do a change of variables with $v = 3u/\lp^2$ to get

\begin{equation}
	\int\dd{u}\psi_\Lambda^*\psi_{\Lambda'} = \frac{2\pi\lp^2\Lambda^2}{3}\delta(\Lambda - \Lambda'),
\end{equation}
wherein the last step, we made use of the following property of the $\delta$-function

\begin{equation}
	\delta(f(x)) = \frac{\delta(x - x_0)}{|f'(x)|},
\end{equation}
where $f(x_0) = 0$. The immediate implication of this is that in minisuperspace, the Kodama state is orthogonal under different choices of cosmological constant. From this statement, we can see that the norm of the state

\begin{equation}
    \Psi[A, T] = \int\dd\Lambda\,\amp(\Lambda)\,e^{i\Lambda T/\lp^2}\psi_\Lambda[A],
\end{equation}
again in the minisuperspace limit becomes

\begin{equation}
    \int\dd{u}|\Psi(u, T)|^2 = \frac{2\pi\lp^2}{3}\int\Lambda^2\dd\Lambda\,|\amp(\Lambda)|^2,
\end{equation}
which is independent of $T$. Thus, the evolution of the state with respect to the unimodular time is unitary\footnote{This follows because unitary operators preserve norms, so if the norm is time independent that any unitary operator will trivially preserve that.}, consistent with our interpretation of this wave function being a stationary state of the unimodular Hamiltonian. This is the connection representation of the unimodular wave packets that were found in~\cite{Alexandre22, Alexandre23a, Alexandre23b}.

This unitary wave-packet description makes the Vilenkin tunneling boundary condition especially interesting. For a real spectral parameter, a self-adjoint Robin condition at the origin enforces vanishing probability current , whereas a purely outgoing stationary wave carries nonzero flux at large scale factor. The two conditions cannot hold simultaneously for a nontrivial real-energy stationary solution. We now show that the outgoing Vilenkin wave can instead be understood as a generalized resonance mode, associated with a complex pole of the scattering problem, while the underlying real-energy wave packets remain unitary.

\section{The Vilenkin Wave Function In Unimodular Time}\label{sec:vilenkin_res}

The covariant Schr\"odinger equation framework offers a different interpretation of the Vilenkin wavefunction. In ordinary scattering, a resonance is a state temporarily confined behind a barrier that eventually leaks into an outgoing channel. A rapid change in the scattering phase shift identifies its real-energy location; a pole of the continued outgoing amplitude determines its  lifetime. Unimodular time makes this distinction meaningful for a cosmological wavefunction.

For a closed de Sitter minisuperspace with $q = a^2$, the Wheeler--DeWitt equation is
\begin{equation}
    \qty[-\qty(\frac{\lp^2}{3\vol_c})^2 \dv{^2}{q^2} + 1 - \frac{\Lambda}{3}q]\psi_\Lambda(q) = 0.
\end{equation}
This can be expressed as an eigenvalue problem by the following

\begin{equation}
    \hat h\psi_\Lambda = -\Lambda\psi_\Lambda,\qquad \hat{h} = \frac{3}{q}\qty[\qty(\frac{\lp^2}{3\vol_c})^2\dv{^2}{q^2} - 1].
\end{equation}
With the dimensionless time normalized so that $\cep\tau = \Lambda T/\lp^2$, define
\begin{equation}
    s = \frac{3\vol_c}{\lp^2}q,\qquad
    \cep = \frac{\lp^2\Lambda}{9\vol_c},\qquad
    \tau = \frac{9\vol_c}{\lp^4}T.
\end{equation}
The dimensionless Hamiltonian and its time-dependent equation become
\begin{equation}
    \hat h(s) = \frac{1}{s}\qty(\dv{^2}{s^2} - 1),
\end{equation}
\begin{equation}
    i\pdv{\Psi}{\tau} = \hat h(s)\Psi.
\end{equation}
Thus $\Psi_\cep(s,\tau) = e^{i\cep\tau}\phi_\cep(s)$ obeys
\begin{equation}
    \qty[-\dv{^2}{s^2} + 1 - \cep s]\phi_\cep(s) = 0.
\end{equation}
Writing $z = \cep^{-2/3}(1 - \cep s)$ gives the Airy equation
\begin{equation}
    \dv[2]{\phi_\cep}{z} = z\phi_\cep,\qquad
    \phi_\cep(z) = \mu_\cep\ai(z) + \nu_\cep\bi(z).
\end{equation}

A fixed real Robin condition at $s = 0$,
\begin{equation}\label{eq:robin}
    \beta\phi_\cep(0) + \phi_\cep'(0) = 0,
\end{equation}
where $\beta \in \realn$, defines a self-adjoint boundary condition for the real-energy problem with measure $s\,\dd{s}$. The classically allowed region is $s > 1/\cep$. Its WKB phase is
\begin{equation}
    \theta(s,\cep)=\int_{1/\cep}^{s}\sqrt{\cep x-1}\,\dd{x}
    = \frac{2}{3\cep}(\cep s-1)^{3/2}.
\end{equation}
To leading WKB order $\ai(z)$ and $\bi(z)$ give sine and cosine waves with phase $\theta + \pi/4$. The particular combination
\begin{equation}\label{eq:vilenkinoutgoing}
    g_V(s;\cep) = \ai(z) + i\bi(z)
\end{equation}
is a purely outgoing branch in the Vilenkin convention. At real $\cep$, this  traveling wave does not by itself satisfy a generic real Robin condition: a self-adjoint scattering state contains both incoming and outgoing waves.

For real $\cep$, write that scattering state as
\begin{equation}
    \phi_\cep(s)=\cep^{-1/6}
    \qty[\cos\delta(\cep)\,\ai(z) + \sin\delta(\cep)\,\bi(z)],
\end{equation}
with $z_0 = \cep^{-2/3}$. The Robin condition gives
\begin{equation}\label{eq:phaseshift}
    \tan\delta(\cep) = -\frac{D_A(\cep)}{D_B(\cep)},
\end{equation}
where
\begin{align}
 D_A(\cep)& = \cep^{1/3}\ai'(z_0) - \beta\ai(z_0),\\
 D_B(\cep)& = \cep^{1/3}\bi'(z_0) - \beta\bi(z_0).
\end{align}
Equivalently, introducing $L_A=\cep^{1/3}\ai'(z_0)/\ai(z_0)$ and $L_B=\cep^{1/3}\bi'(z_0)/\bi(z_0)$,
\begin{equation}
    \tan\delta(\cep)=-\frac{\ai(z_0)}{\bi(z_0)}
    \frac{L_A(\cep)-\beta}{L_B(\cep)-\beta}.
\end{equation}
A narrow resonance is signaled when the denominator vanishes at a real $\cep_R$,
\begin{equation}\label{eq:rescenter}
    \beta = L_B(\cep_R).
\end{equation}
Indeed, expanding $D_B$ near $\cep_R$ gives the familiar rapid phase variation
\begin{equation}\label{eq:BWphase}
    \tan\delta(\cep)=-\frac{\gamma_{G}}{\cep - \cep_R} + \order{1},
\end{equation}
where $\gamma_{G}$ is the positive narrow-resonance scale defined below. This phase-shift crossing identifies the resonance center, but not yet the Gamow state itself.

Why must we continue to complex $\cep$? At real $\cep$, a real Robin condition forces the probability current at $s = 0$ to vanish. The current is conserved, so a stationary real-energy solution cannot carry a nonzero \emph{purely outgoing} flux at large $s$. The physical scattering state must contain both directions. A Gamow state instead imposes the outgoing condition alone; it is not an eigenstate of the self-adjoint real-energy problem. We therefore continue the spectral parameter into the complex plane and demand that the \emph{same} Robin condition hold for the outgoing Vilenkin solution. This yields the pole equation
\begin{equation}\label{eq:gamowcondition}
    D_A(\cep_*) + iD_B(\cep_*)=0.
\end{equation}
For a narrow resonance, expand about the real center defined by Eq.~\eqref{eq:rescenter}. Since $D_B(\cep_R) = 0$, the leading pole location is
\begin{equation}
    \eval{\cep_*}_{\rm leading} = \cep_R + i\frac{D_A(\cep_R)}{D_B'(\cep_R)}.
\end{equation}
For $0 < \cep_R \ll 1$, the Airy asymptotics give the leading pole position
\begin{align}\label{eq:width}
    \cep_*& = \cep_R + i\gamma_{G},\\
    \gamma_{G}& = 4\exp\qty[-\frac{4}{3\cep_R}]
    \qty[1 + \order{\cep_R}].\nonumber
\end{align}
Here $\gamma_{G} = \operatorname{Im}\cep_*$ is the positive \emph{amplitude} decay rate per unit dimensionless unimodular time $\tau$; the probability decay rate is $\Gamma = 2\gamma_{G}$.  The exponent is twice the under-barrier action, $2\int_0^{1/\cep_R}\sqrt{1 - \cep_Rs}\,\dd{s} = 4/(3\cep_R)$. Using the definition of $\cep$ above, we find
\begin{align}\label{eq:physicalgamma}
    \frac{4}{3\cep_R} &= \frac{12\vol_c}{\lp^2\Lambda_R},\\
    \gamma_{G} &= 4\exp\qty[-\frac{12\vol_c}{\lp^2\Lambda_R}]
    \qty[1 + \order{\cep_R}].\nonumber
\end{align}
The real resonance value, $\Lambda_R$, is fixed by the Robin condition $\beta = L_B(\cep_R)$, not chosen independently. The corresponding amplitude decay rate per unit four-volume time $T$ is
\begin{equation}
    \gamma_T = \frac{9\vol_c}{\lp^4}\gamma_{G}.
\end{equation}
Thus, a smaller positive $\Lambda_R$ produces exponentially slower leakage through the Wheeler--DeWitt barrier.

The corresponding \emph{resonance contribution}, rather than the complete physical wavefunction, is
\begin{equation}\label{eq:gamowtime}
    \Psi_{G}(s,\tau) = \mathcal{N}_{G} e^{i\cep_R\tau}e^{-\gamma_{G}\tau} \qty[\ai(z_*) + i\bi(z_*)],
\end{equation}
with $z_* = \cep_*^{-2/3}(1-\cep_*s)$. Its outgoing spatial dependence and exponential decay in positive unimodular time are the characteristic Gamow properties. The complex $\cep_*$ is a resonance pole of the analytically continued spectral problem, not a complex eigenvalue of the self-adjoint physical Hamiltonian. Its non-normalizability therefore does not conflict with unitary evolution of superpositions of real-energy scattering states. Whether this pole contributes to a particular physical wave packet depends on the packet and its analytic continuation.

\section{Discussion}\label{sec:discussion}

Using the Ashtekar variables with the self-dual restriction, we have constructed a background independent wave function, $\Psi[A, T]$ that evolves with respect to the volume of the spacetime manifold, $\mcal$, which acts as a gauge invariant clock. The spacetime manifold's volume acts as a ``good clock" parametrized by the global time variable $T[\Sigma]$, where $\Sigma$ is the spacelike hypersurface which we call the unimodular time. A ``good clock" being a function that satisfies the following criteria: (1) it is decoupled from all other dynamical variables of the Hamiltonian and (2) the wave function, $\Psi[A, T]$ that depends on this global time variable is tightly peaked around some monotonic function of the global time variable. We have also demonstrated that by constructing the general solution of the unimodular Hamiltonian off of the back of the Kodama state, we are able to identify the Kodama state, indexed by the cosmological constant $\Lambda$, as a stationary state of quantum gravity which, when reduced to minisuperspace and subject to the naive inner product, is delta function normalizable. We have also generalized old ideas in the literature of the Chern-Simons functional being used as a global time variable~\cite{Smolin95, Magueijo18, Magueijo21b}. In these previous works as well as many, the Kodama state was reduced to the minisuperspace limit~\cite{Magueijo20a, Magueijo20b, Magueijo21a, Magueijo21b, Alexandre22, Alexandre23a, Alexandre23b} and then shown to be a natural clock for quantum gravity. In this work, we show the full \emph{background-independent} solution, itself ultimately can be used as a clock. Finally, within the unimodular framework, we were able to show that the Vilenkin wave function can be viewed as a gravitational Gamow resonance that lives in the rigged Hilbert space of the Wheeler DeWitt equation. 

The main conceptual result is therefore broader than the minisuperspace resonance. The covariant unimodular clock provides a time-dependent Wheeler--DeWitt equation with exact, background-independent solutions constructed from a time-dependent extension of the  Kodama state. The Kodama functional retains its role as a stationary  solution, but the  four-volume time permits time-evolving superpositions. This addresses the frozen-state formulation of the problem of time without requiring a symmetry reduction. The explicit norm calculation and the Gamow analysis, however, are established  in minisuperspace, and should not be mistaken for a proof of unitarity or resonance behavior on  the full superspace.

The resonance interpretation also changes what one should ask of the wavefunction of the Universe. A normalizable state prepared with a self-adjoint boundary condition and a generalized outgoing Gamow mode are not competing definitions of the same Hilbert-space vector. The first follows  unitary quantum evolution; the second characterizes the analytic continuation of its scattering spectrum and may describe transient leakage into an expanding geometry. In this sense, our construction offers a concrete interpretation of the non-normalizable Vilenkin outgoing solution once the wave evolves according to the unimodular time . It does not by itself identify a unique cosmological state-preparation prescription or establish equivalence with the Hartle--Hawking no-boundary proposal. In particular, the path-integral contour and the stability of inhomogeneous perturbations remain separate questions. A sharper test of this picture would be to construct a physical wave packet whose analytically continued amplitude has a nonzero residue at the Vilenkin pole and to study the corresponding fluctuation state.


Although we are presenting the unimodular time as a global, covariant, and gauge invariant (up to an overall constant) clock, it is still fair to ask how to interpret such an object. We showed that Eq~\eqref{eq:manifold_clock}, the unimodular time is exactly determined by the volume of the total manifold. To get a sense of how to think about this notion, recall that when working in minisuperspace the wave packets as a function of the unimodular time is tightly peaked around the Chern-Simons time $\tcs = a^3\vol_c/(3H)$, where $\vol_c$ is the comoving Hubble volume and $H$ is the Hubble parameter during inflation. One very literal interpretation of the unimodular time is that it represents a measurement of the comoving Hubble volume integrated up to some redshift. And therefore, the more precisely one can measure the value of the cosmological constant, the longer the cosmic history that a comoving observer is able to probe.

Another interpretation harkens back to using the Chern-Simons functional as a clock. Given that the age of the Universe is given by $H^{-1}$, $\tcs$ has the form of $\Vol(\mcal)$. Because the cosmological constant is canonical partners with the unimodular time, there is a uncertainty relationship between the variables $\Lambda$ and $T$. Thus, a measurement of the value of $\Lambda$ necessarily corresponds to losing information of $\Vol(\mcal)$. Since the comoving Hubble volume is an observable, the uncertainty must be in the age of the Universe. Given that we are considering a model of a closed Universe, we may interpret the unimodular time as the number of cycles of expansion and contraction that the Universe has undergone as initially proposed in~\cite{Magueijo18}.

Some things that bares clarification is that we have not solved the cosmological constant problem~\cite{Padilla15}, nor have we solved the problem of defining local, gauge invariant, observables in quantum gravity~\cite{Harlow26}. We are not able to address the critiques of the unimodular gravity program for the problem of time raised in~\cite{Kuchar91, Isham92}. The problem with the approach of unimodular gravity that was raised by~\cite{Kuchar91} can be traced to the idea that the global time variable doesn't pick out a particular spacelike hypersurface, $\Sigma$. As a result, causality cannot be guaranteed in this construction. What we have done is build off of the work of~\cite{Smolin09, Smolin11, Magueijo18} and show that the Kodama state can be used as a background independent candidate for defining the stationary solution for the wave function of the Universe. By centering the role that the Kodama state plays, we hope to inspire future generations to take the existence of this term seriously. We speculate that perhaps the fact that the Chern-Simons functional has these large gauge transformations resulting in a shift $\ycs$ by $\ycs\rightarrow\ycs + (4\pi)^2n$ could play a role in picking out a particular topology in order to regulate this term.

\begin{acknowledgments}

We thank Cyril Creque-Sarbinowski, Robert Brandenberger, Tatsuya Daniel, Lawrence Edmond IV, and Philip Phillips for useful discussions and comments.

\end{acknowledgments}

\appendix






\section{The Kodama State Satisfies All Constraints}\label{app:kodama_constraints}

Now we have the Hamiltonian (constraint)

\begin{equation}
    \ham = \ep_{ijk}E^{ai}E^{bj}\qty(F^k_{ab} + \frac{\Lambda}{3}\ep_{abc}E^{ck}) = 0,
\end{equation}
which acts on some wave function $\psi[A]$. Applying the regular canonical quantization procedure i.e.

\begin{equation}
    \hat{E}^{ai}\rightarrow \lp^2\fdv{A_{ai}},
\end{equation}
the annihilation of the quantum state becomes

\begin{equation}
    \hat{\ham}\psi[A] = \lp^4\ep_{ijk}\fdv{A_{ai}}\fdv{A_{bj}}\qty(F^k_{ab} - \frac{\lp^2\Lambda}{3}\ep_{abc}\fdv{A_{ck}})\psi[A] = 0.
\end{equation}
For this particular operator ordering, the form of the state $\psi$ that satisfies this annihilation is most easily solved by assuming that the term in brackets vanishes (the so-called self-dual constraint)

\begin{equation}
    \ep_{abc}\fdv{\psi}{A_{ck}} = \frac{3}{\lp^2\Lambda}F^k_{ab}\psi[A].
\end{equation}
Contracting both sides with $\ep^{dab}$ gives us

\begin{equation}
    \fdv{\psi}{A_{ai}} = \frac{3}{2\lp^2\Lambda}\ep^{abc}F^i_{bc}\psi[A].
\end{equation}
Recognizing the term that is multiplying the wave function, we can write

\begin{equation}
    \fdv{\psi}{A_{ai}} = -\frac{3}{2\lp^2\Lambda}\fdv{\ycs}{A_{ai}}\psi[A].
\end{equation}
From here it is easy to see what the solution has to be. Since we're getting something proportional to the original function multiplying the derivative of some other function, we can write

\begin{equation}
    \psi_\Lambda[A] = \mathcal{N}\exp(-\frac{3}{2\lp^2\Lambda}\ycs),
\end{equation}
where $\mathcal{N}$ is some normalization constant. We can also see that the Kodama state is simultaneously annihilated by the other two constraints. The classical constraints are

\begin{equation}
	\ep_{abc}B^{bk}E^c_k = 0, \quad\cd_a E^a_i = 0,
\end{equation}
where $B^{ai} = -\ep^{abc}F^i_{bc}$. Let us check the claim that the Kodama state does in fact satisfy the other constraints:

\begin{equation}
	\ep_{abc}\hat{B}^{b}_k\hat{E}^{ck}\psi_\Lambda[A] = -\frac{3}{2\lp^2\Lambda}\ep_{abc}B^b_kB^{ck}\psi_\Lambda\Lambda[A] = 0,
\end{equation}
where we used the fact that $B^b_kB^{ck}$ is symmetric in $b,c$ and $\ep_{abc}$ is antisymmetric in $b,c$. Next we have

\begin{equation}
	\begin{split}
	    \cd_a\hat{E}^{ai}\psi_\Lambda[A] &= -\frac{3}{2\lp^2\Lambda}\cd_a(B^{ai}\psi_\Lambda[A])\\ &= -\frac{3}{2\lp^2\Lambda}B^{ai}\cd_a\psi_\Lambda[A]\\ &= -\frac{3}{2\lp^2\Lambda}B^{ai}\p_a\psi_\Lambda[A] = 0.
	\end{split}
\end{equation}
\newpage

\bibliographystyle{apsrev4-2}
\bibliography{unimodular_kodama_state_time}

\end{document}